\documentclass[a4paper]{jpconf}
\usepackage{graphicx}
\begin{document}
\title{Hydrodynamic Overview at Hot Quarks 2016}

\author{Jacquelyn Noronha-Hostler}

\address{\small{\it  Department of Physics, University of Houston, Houston, TX 77204, USA}}

\ead{jakinoronhahostler@gmail.com}

\begin{abstract}
This presents an overview of relativistic hydrodynamic modeling in heavy-ion collisions prepared for Hot Quarks 2016, at South Padre Island, TX, USA. The influence of the initial state and viscosity on various experimental observables are discussed. Specific problems that arise in the hydrodynamical modeling at the Beam Energy Scan are briefly discussed.  

\end{abstract}

\section{Introduction}
The Quark-Gluon Plasma (QGP) existed microseconds after the Big Bang where its size was significantly larger and its expansion significantly slower than the QGP experimentally measured in heavy ion collisions.  The QGP produced in the laboratory is the hottest, smallest, and densest fluid known to mankind so the techniques used to describe its properties are still being developed. With the discovery that the QGP formed at RHIC and the LHC was a nearly perfect liquid, a ``standard model" of heavy-ion collisions has emerged that includes fluctuating initial conditions, event-by-event relativistic viscous hydrodynamics, and a hadronic afterburner.  

The main signatures of nearly perfect fluidity are the flow harmonics (Fourier coefficients of the particle spectra) that can be reproduced within relativistic hydrodynamical calculations with an extremely small shear viscosity to entropy density ratio, $\eta/s\sim 0.08$.  Originally, it was understood that there should be a lower bound for $\eta/s\sim 1/4\pi$ by combining a quasiparticle description with the uncertainty principle \cite{Danielewicz:1984ww}. The derivation of the KSS limit from strong coupling holography \cite{Kovtun:2003wp} initially gave support to such a bound, although, now it is known that there are holographic examples where $\eta/s$ can be even smaller, e.g., \cite{Kats:2007mq,Brigante:2007nu,Brigante:2008gz,Buchel:2008vz,Critelli:2014kra,Finazzo:2016mhm}. It is expected that a minimum exists close to the crossover temperature \cite{Aoki:2006we} as one goes from a strongly interacting QGP phase into an eventually weakly interacting hadron gas phase  \cite{NoronhaHostler:2008ju,NoronhaHostler:2012ug}.  Further references for the temperature dependence of viscosity can be found in \cite{Noronha-Hostler:2015qmd}.

While a reasonable estimate for the range of values of $\eta/s$ can be made using sophisticated Bayesian techniques \cite{Bernhard:2015hxa,Bernhard:2016tnd}, its exact value is extremely dependent on the initial state formed immediately after the two heavy-ions collide.  Over the years, a plethora of initial state models have been developed, each of which has a corresponding range of valid transport coefficients that allow for reasonable fit to experimental data.  One of the most pressing issues remaining in heavy ions is finding observables either sensitive to only the initial state or only to transport coefficients, which will be discussed in detail in this proceedings.

\section{Collective Flow}

When two heavy ions are smashed together, clear geometrical effects occur depending on if they hit head-on (central collisions), have a grazing collision (peripheral), or hit somewhere in between (mid-central).  Experimentally, heavy ions are collided billions of times (each collision is an event) and each event produces a different number of particles that participated in the collision, known as multiplicity.  The more central the collisions, the larger multiplicities are produced so the events are then sorted by their multiplicities into central classes where $0\%$ centrality  have the highest multiplicities and $100\%$ centrality indicates the lowest possible multiplicities. Central collisions produce on average a circular shape in the transverse plane to the beam axis whereas as an approximate almond shape is produced for mid-central collisions and beyond.  

Due to quantum fluctuations in the initial position of protons and neutrons within each ion, a multitude of shapes can be produced \cite{Alver:2010gr}.  Each highly inhomogeneous initial condition runs separately through hydrodynamics on an event-by-event basis. Experimentally the initial state cannot be measured directly, rather pressure gradients convert the initial geometrical shapes into a corresponding momentum space anisotropy, measured via flow harmonics.  To obtain the flow harmonics, one calculates the Fourier coefficients of the particle spectra (with special care  to reproduce the exact way experimentalists measure flow harmonics \cite{Luzum:2012da} where multiplicity weighing and centrality rebinning should not be ignored \cite{Gardim:2016nrr,Betz:2016ayq}). 

A number of parameters that go into hydrodynamical modeling such as the initial time after which one assumes the system admits a hydrodynamic description, and also the switching temperature below which a hadronic transport is used.  The initial time, $\tau_0$, depends on the collisional energy as well as the initial condition type and these issues were discussed in more detail in the Beam Energy Scan and anisotropic hydrodynamics talks at this conference. The maximum switching temperature, $T_{SW}$, is constrained by the hadronization temperature indicated from Lattice QCD \cite{Borsanyi:2014ewa} to ensure that one switches to the correct degrees of freedom.  

\begin{figure}[h]
\centering
\includegraphics[width=20pc]{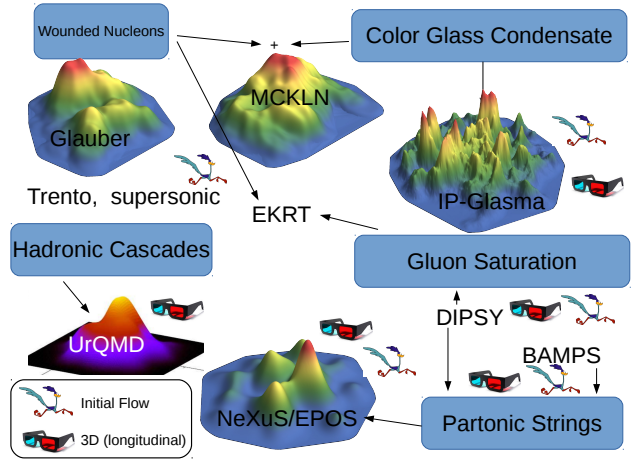}
\caption{\label{fig:IC}Initial conditions used in heavy ion collisions organized by their basic assumptions.}
\end{figure}

The flow harmonics themselves are most sensitive to the choice in the initial conditions as well as the transport coefficients ($\eta/s$, bulk viscosity to entropy density, $\zeta/s$, and their corresponding relaxation times- $\tau_{\pi}$ and $\tau_{\Pi}$, respectively). In the next two sections experimental observables that constrain the initial state and transport coefficients are discussed. 

\section{Separating the Initial State from Transport Coefficients}

At LHC energies a number of initial conditions such as IP-Glasma \cite{Gale:2012rq}, EKRT \cite{Niemi:2015qia}, and Trento (tuned to IP-Glasma) \cite{Moreland:2014oya} manage to fit well the two and four particle flow cumulants as well as the distribution of $v_n$'s. In Fig.\ \ref{fig:IC} a schematic cartoon of the most well-known initial condition models categorized by their basic properties  is shown.  Additionally, extremely accurate predictions for the flow harmonics of LHC run 2 to the order of $\sim 5\%$ \cite{Niemi:2015voa,Noronha-Hostler:2015uye} were made and later experimentally confirmed in \cite{Adam:2016izf}. However, the different models used in these predictions varied parameters like viscosity, freeze-out, and the inclusion of initial flow. 

It has been well-established that an approximately linear relationship exists between the initial energy/entropy density eccentricities, $\varepsilon_n$, and the experimentally measured flow harmonics $v_n$'s \cite{Teaney:2010vd,Gardim:2011xv,Teaney:2012ke,Niemi:2012aj,Gardim:2014tya} and that sub-nucleon fluctuations do not appear to play a significant role in the calculation of the lowest order harmonics in large collision systems \cite{Noronha-Hostler:2015coa}. That being said, higher order flow harmonics are significantly more complicated and depend on a variety of eccentricities \cite{Gardim:2011xv,Gardim:2014tya}. Additionally, $v_1$ is especially complicated \cite{Gardim:2014tya} and may depend on the full $T^{\mu\nu}$ initialization \cite{Gardim:2011qn}. 

\begin{figure}[h]
\centering
\includegraphics[width=20pc]{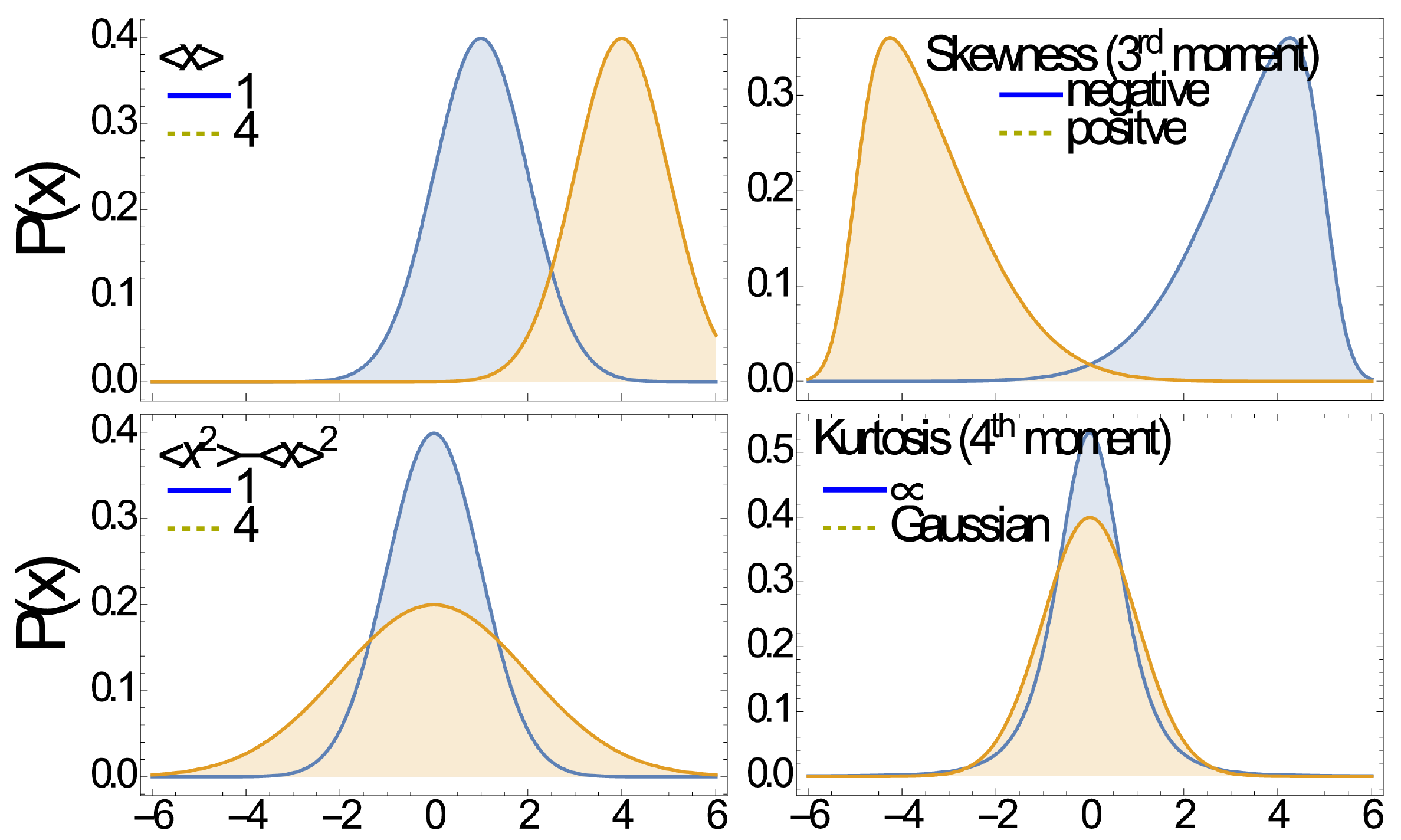}
\caption{\label{fig:mom}First moment (mean), second moment (variance), third moment (skewness), and fourth moment (kurtosis) of a distribution.}
\end{figure}

In order to more easily quantify the distribution of flow harmonics, multiparticle cumulants are used.  Cumulants of the $v_2$ distribution are directly connected to the moments of the distribution  via $v_2\{4\}/v_2\{2\})^4=2-\langle v_2^4\rangle/\langle v_2^2\rangle^2$, which can indicate the degree of which the system is fluctuating (see Fig.\ \ref{fig:mom}).  If there are no fluctuations in the system the $p_T$-integrated $v_2\{4\}/v_2\{2\}\rightarrow 1$, whereas $v_2\{4\}/v_2\{2\}< 1$ and $v_2\{4\}\sim v_2\{6\} \sim \dots$ is a sign of the collective behavior measured in heavy ion collisions.  Note, however, that the higher order cumulants are not exactly identical, small deviations can exist due to the skewness of the initial conditions \cite{Giacalone:2016eyu}.  Finally, complications do exist for more peripheral collisions where deviations are seen between the linear mapping of the initial eccentricities and the final elliptical flow \cite{Niemi:2015qia}, which can be explained due to cubic response \cite{Noronha-Hostler:2015dbi}.

Recently, symmetric cumulants \cite{ALICE:2016kpq} that measure the correlation of different order flow harmonics on an event-by-event basis have been measured in PbPb collisions and $SC(3,2)$ (which involves elliptic and triangular flow) appears to be almost entirely driven by the initial eccentricities.  While it was thought that $SC(4,2)$ was sensitive to the choice in viscosity, much of that disappears after a proper treatment of multiplicity weighing and centrality binning \cite{Gardim:2016nrr}. 

Another possibility that exists for extracting the properties of the initial state is via the correlations between soft and hard sectors of heavy ion collisions, i.e., soft-hard event engineering (SHEE) as described in \cite{Noronha-Hostler:2016eow} where it was found that the high $p_T$ flow harmonics are directly linked to fluctuating initial eccentricities (due to the path length dependence of jet quenching).  Moreover, it was recently found that viscosity plays little role in the high $p_T$ flow harmonics \cite{Betz:2016ayq} so it's especially interesting to exploit this relationship. Further studies may also be possible in the heavy flavor sector as well \cite{Nahrgang:2014vza}, as done for instance in \cite{Prado:2016xbq}.

Interestingly enough, it appears that there is a direct link between the symmetric cumulants and the event plane correlations \cite{Giacalone:2016afq}, though the latter are significantly more sensitive to viscous effects \cite{Niemi:2015qia}. In fact, event plane correlations appear to be very sensitive to the temperature dependence of $\eta/s$ where a small viscosity in the hadron resonance gas phase seems to be preferred (such a small $\eta/s$ can occur due to a large number of massive, degenerate states close to the crossover transition \cite{NoronhaHostler:2008ju,NoronhaHostler:2012ug}). Additionally, $v_4\{4\}^4$ may indicate that a larger $\eta/s$ is needed but further testing is still needed \cite{Giacalone:2016mdr}.

The extraction of transport coefficients is significantly more intricate due to the interplay between shear and bulk viscosity \cite{Noronha-Hostler:2013gga,Noronha-Hostler:2014dqa,Ryu:2015vwa}.  This is further complicated due to the large influence of the bulk viscous corrections at hadronization. In fact, different implementations of the bulk viscosity correction to the particle distribution at freeze-out can either increase the shear viscosity \cite{Bernhard:2016tnd,Noronha-Hostler:2013gga,Noronha-Hostler:2014dqa} or decrease it, as in \cite{Ryu:2015vwa}. Furthermore, second order transport coefficients may also become relevant \cite{Molnar:2013lta,Finazzo:2014cna} especially in small collision systems, and the effects of many of these coefficients have not been systematically studied within the context of event-by-event hydrodynamics.

Additionally, $\langle p_T\rangle$ was recently used in \cite{Ryu:2015vwa} (and later in \cite{Bernhard:2016tnd}) to help constrain the temperature dependence of the bulk viscosity. There is a very intricate connection between the temperature dependence of $\zeta/s$ and the equation of state (bulk viscosity is only nonzero in a non-conformal system, which indicates that there should be a peak in bulk viscosity at the crossover \cite{Karsch:2007jc}) so this quantity is expected to play an even more significant role in the Beam Energy Scan.

\section{Beam Energy Scan}

So far in this paper we focused on LHC energies where the baryon chemical potential, $\mu_B$, is approximately zero. With the Beam Energy Scan at RHIC one is able to probe matter that would correspond to other regions in the QCD Phase Diagram and, hopefully, also approach the critical end point.  However, a significant number of questions remain at almost all level of hydrodynamical modeling. Initial calculations within kinetic theory/hadron resonance gas \cite{Demir:2008tr,Denicol:2013nua,Kadam:2014cua}, which do not contain a critical point, show a decrease in $\eta/s$ as one increases $\mu_B$. However, if the dynamic universality class of the QCD critical point is that of model H \cite{Son:2004iv}, then one would expect the transport coefficients to diverge at the critical point.  Assuming that the dynamic universality class is H and there is a diverging viscosity, then this would be a nice explanation for the turning off of triangular flow observed at low energies \cite{Adamczyk:2016exq}.  

However, there may be a more simple explanation for the decrease in $v_3$ at low energies that connects to the decreased amount of time spent in the hydrodynamic phase as the beam energy decreases \cite{Auvinen:2013sba}.  This would be more consistent either with no critical point (or perhaps one placed at much larger $\mu_B$) or perhaps with a different dynamic universality class such as that of model B \cite{Rougemont:2015ona}. Significantly more research needs to be done to better understand both the location of the critical point as well as the dynamic properties of the expanding QGP in this regime and, hopefully, the results from the upcoming Beam Energy Scan II run at STAR will lead to unambiguous experimental verification of QCD critical behavior.

\section{Outlook}

Relativistic viscous hydrodynamics has become a powerful tool to explore the strongly interacting QCD matter formed in heavy ion collisions.  Significant progress has been made in finding ``orthogonal" measurements that can independently aid in extracting either the initial state of the system or its transport properties.  A number of open-ended questions still remain such as if there is still perfect fluidity close to the QCD critical point as well as what special considerations need to be taken into account at finite baryon density.  While not covered here, other important questions discussed at this conference remain, e.g., the need for anisotropic hydrodynamics \cite{Nopoush:2014pfa} and also pre-equilibrium flow \cite{Heinz:2015arc,Keegan:2016cpi}. Additionally, the origin of flow harmonics (and their corresponding cumulants) in small systems is still up for debate \cite{prit}.

\section*{Acknowledgements}  JNH thanks the HotQuarks2016 organizers for inviting her to give the Hydrodynamic Overview of Heavy-Ion Collisions  and she was supported by the National Science Foundation under grant
no. PHY-1513864.

\end{document}